\documentclass[letterpaper]{article}
\usepackage{aaai2026}
\usepackage{times}  
\usepackage{helvet}  
\usepackage{courier}  
\usepackage[hyphens]{url} 
\usepackage{graphicx} 
\usepackage{natbib}  
\usepackage{caption} 
\usepackage{amsmath}
\usepackage{amssymb}
\usepackage{mathtools}
\usepackage{physics}
\usepackage{multirow}
\usepackage{xcolor}

\title{Architectural Robustness to Label Noise Using Variational Quantum Circuits for Medical Image Classification}
\author{
    Yasoda Lasiru Ariyarathna Epa\equalcontrib\textsuperscript{\rm 1},
    Yaseema Rusiru Ariyarathna Epa\equalcontrib\textsuperscript{\rm 1},
    Sandaru Liyanaarachchi\equalcontrib\textsuperscript{\rm 1},
    Kavindu Jayathissa\equalcontrib\textsuperscript{\rm 1},
    Kalana G. Abeywardena\textsuperscript{\rm 2},
    Kithmin Wickremasinghe\textsuperscript{\rm 3},
    Theshani Nuradha Piliththuwasam Gallage\textsuperscript{\rm 4},
    Chamira U. S. Edussooriya\textsuperscript{\rm 1}
}

\affiliations{
    \textsuperscript{\rm 1}University of Moratuwa, Sri Lanka \\ 
    \textsuperscript{\rm 2}University of Toronto, Canada \\
    \textsuperscript{\rm 3}University of British Columbia, Canada \\
    \textsuperscript{\rm 4}University of Illinois Urbana-Champaign, USA
}

\begin{document}

\maketitle

\begin{abstract}
Label noise is prevalent in medical image datasets, arising from observer disagreement and ambiguous cases and deep networks memorize the mislabeled examples at the cost of generalization. Current solutions such as loss correction and sample selection add training complexity and generally assume the noise rate is known or estimable. This paper investigates whether robustness can be obtained architecturally, by routing a frozen backbone's features through a low-dimensional bounded bottleneck and classifying them with a head whose function class is intrinsically smooth. Variational quantum circuits with angle encoding and a linear cost function realise exactly such a class.
Their use in medical imaging, however, has been limited by the small number of qubits available on noisy intermediate-scale devices, which typically restricts inputs to low-resolution grayscale images. We therefore couple a frozen ImageNet-pretrained ViT-B/16 backbone to a trainable linear bottleneck and an eight-qubit variational quantum circuit, allowing $224\times224$ RGB inputs to be processed. Under clean labels the model leads five of six MedMNISTv2 benchmarks against published hybrid quantum-classical models, with margins of $17.4$ and $15.7$ percentages on BloodMNIST and PathMNIST. The advantage emerges under corrupted labels. At symmetric noise rates of $30\%$ and $50\%$ the model outperforms many published quantum and noise-robust classical baselines on all six datasets without any explicit noise-handling component. Against a control that shares the backbone, bottleneck, parameter count and optimizer and replaces only the circuit with a ReLU layer, it retains $89.75\%$ accuracy on PathMNIST at $50\%$ noise where the control falls to $78.96\%$. Direct measurement of training dynamics shows that this advantage is not an inability to fit corrupted labels, which the circuit retains, but greater stability under prolonged training on them.
\end{abstract}

\section{Introduction}
Label noise is an intrinsic property of almost all medical image datasets. Because diagnostic categories exist along continuous visual spectra, expert classification is inherently variable, leaving borderline cases without definitive labels and allowing automated pipelines to propagate systematic errors at scale \cite{LITJENS201760,ma2025lnmbench}. Reported disagreement rates in histopathology and dermatology routinely reach double digits and the errors concentrate on the visually ambiguous cases that matter most clinically.

Deep networks are unusually vulnerable to this corruption, since networks with sufficient capacity can interpolate arbitrary label assignments, including entirely random ones \cite{zhang2017}. After an initial phase in which the true class structure is learned, prolonged training continues to reduce the loss by memorizing individual mislabelled examples and test accuracy degrades. Established countermeasures operate on the training procedure rather than the model, loss-correction methods reweight the objective, sample-selection methods discard suspected corrupted examples and transition-matrix methods estimate the noise process explicitly \cite{duenet,diaz2021,ma2025lnmbench}. All add training complexity, most introduce hyperparameters tuned per noise level and many require the noise rate to be known or reliably estimated, an assumption difficult to satisfy clinically, where the true labels are by definition unavailable.

This motivates the question our paper addresses, can robustness be built into the architecture instead, by choosing a classifier that cannot easily represent the corruption? Label flips assign conflicting targets to examples that lie close together in feature space, so fitting them requires a decision function that changes sharply between neighbouring points. A classifier restricted to smooth, slowly varying functions is correspondingly limited in how well it can accommodate them, while remaining able to represent the broad boundaries that genuine class structure induces.

Quantum machine learning (QML) offers a natural source of such restricted classifiers. Quantum computers process information in high-dimensional Hilbert spaces and a common QML strategy is to encode classical data into few number of qubits, evolve that state through a circuit whose rotation angles are trainable and read out measurement statistics as features for a downstream classifier. Models of this kind, known as variational quantum circuits (VQCs), are attractive because they represent data compactly \cite{Havl_ek_2019}. Less commonly noted is that the way data enters the circuit places a hard boundary on the complexity of the functions the circuit can express. When each input feature is encoded once, as a single rotation angle, the resulting measurement outputs vary smoothly and slowly with that feature \cite{schuld2021}.  
What is usually treated as a limitation on expressivity is, for our purposes, exactly the property we want.

Deploying such a circuit on clinically realistic images is not straightforward. Quantum neural networks in medical imaging have been constrained by the ``qubit bottleneck'' of noisy intermediate-scale quantum (NISQ) devices \cite{hu2025amplitudeencodingbasedclassicalquantumtransferlearning,Kiwit_2025,Preskill_2018}. Quantum convolutional networks \cite{qcnn2022bokhan} and adaptive hybrid models \cite{ahqcnn2023ajlouni,hqnet2024mazher} demonstrate feasibility but are largely restricted to low-resolution ($28\times28$) grayscale inputs \cite{qcnn2022bokhan,rahman2025nqnn}, discarding fine textural detail on which histological and dermatological diagnosis depends. As the class count grows, fully parameterized circuits also encounter the barren plateau problem, which refers to the exponential decay of gradient magnitudes with circuit width and depth and becomes unstable to train \cite{qtnet2024dhara}. Existing work that addresses label noise in a quantum setting does so with explicit noise-aware components such as error-mitigation routines or purpose-built circuit designs \cite{rahman2025nqnn}, leaving the question of intrinsic architectural robustness open.

We therefore construct a sequential classical-quantum-classical architecture in which a frozen ViT-B/16 backbone performs feature extraction, a trainable linear layer compresses its $768$-dimensional output to an eight-dimensional bounded code and an eight-qubit variational circuit performs the decision-making. We refer to the compression layer and the low-dimensional output it produces as the \emph{bottleneck} throughout. Because the backbone absorbs the image and the circuit sees only the eight-dimensional input, input resolution is decoupled from qubit count and $224 \times 224 \times 3$ inputs are processed with eight qubits.

We make the following contributions.
\begin{itemize}
    \item We characterise the function class of a single-layer encoding VQC, showing that its outputs are bounded trigonometric polynomials with frequency support confined to one harmonic per input coordinate and a Jacobian bounded independently of the trained parameters
(Section~\ref{sec:theory}).

    \item We show that the architecture requires no explicit noise-handling mechanism to outperform %\tn{many of the? DONE} 
    many published quantum and noise-robust classical baselines at $30\%$ and $50\%$ symmetric noise on all six MedMNISTv2 multi-class benchmarks and that it remains competitive under instance-dependent noise (Section~\ref{sec:noise}).

    \item We isolate the circuit's contribution with a control that shares the backbone, bottleneck, parameter count and optimizer and replaces only the $8\rightarrow16$ map with a ReLU layer. The two are indistinguishable on clean labels, but the gap widens with the noise rate to $10.8$ percentage points on PathMNIST at $50\%$ noise (Section~\ref{sec:isolation}).

    \item We measure memorization directly rather than inferring it from test accuracy, tracking each head's agreement with the corrupted labels it was trained on. This shows that the quantum head's contribution is stability under prolonged training rather than an inability to fit corruption (Section~\ref{sec:dynamics}). We do not claim this advantage is intrinsically quantum rather than a consequence of the encoding’s restricted spectral complexity; isolating that question against smoothness-matched classical heads is the direct extension of this work, discussed in Section~\ref{sec:limitations}.
\end{itemize}

\section{Proposed Architecture}
\label{sec:2}
\begin{figure*}[t]
    \centering
    \includegraphics[width=0.75\textwidth]{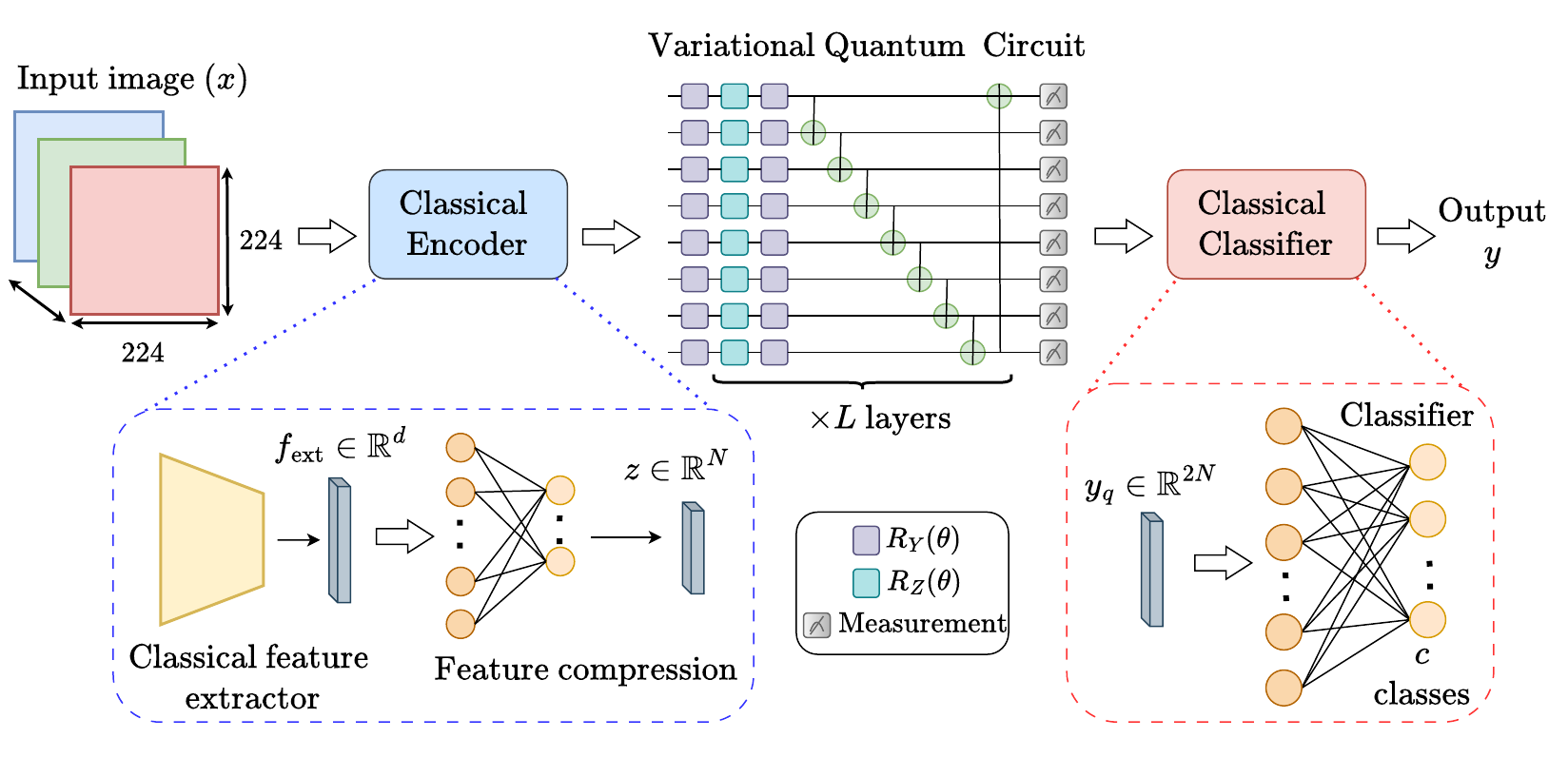}
    \caption{Overview of the proposed hybrid quantum-classical architecture. High-resolution RGB medical images are mapped to a high-dimensional feature vector $f_{\rm ext} \in \mathbb{R}^d$ via a feature extractor $\phi(\cdot)$. It is then compressed into $z \in \mathbb{R}^N$ and mapped to an $N$-qubit VQC. Measurement outcomes $y_q \in \mathbb{R}^{2N}$ are then fed to a classical classifier to get class probability vector $y \in \mathbb{R}^c$.}
    \label{fig:architecture}
\end{figure*}

In this section, we present our proposed hybrid quantum-classical transfer learning framework for medical image classification. The sequential classical-quantum-classical architecture uses a classical feature extractor to produce a high-dimensional representation, which is then projected to a VQC for low-dimensional representation learning. An overview of the architecture is illustrated in Fig.~\ref{fig:architecture}.

\subsection{Classical Feature Extractor}

Current NISQ-era quantum devices are limited by a small number of qubits, making the direct processing of high-resolution RGB medical images (e.g., $224 \times 224 \times 3$) computationally infeasible. To address this, we first employ a classical feature extractor to map the input image to a compact, information-rich latent representation before quantum processing. A deep neural network (DNN) pre-trained on ImageNet \cite{deng2009imagenet} is used as the feature extractor, taking all layers up to but excluding the classification head. This truncated network is the \emph{backbone}, denoted $\phi(\cdot)$. We select ViT-B/16 because, unlike convolutional neural networks (CNNs) with locally constrained receptive fields, its global self-attention \cite{manzari2023medvit} captures the long-range dependencies present in pathological and anatomical structure, such as irregular tumour boundaries and spatially distributed organs \cite{chen2021transunet}.

For an input image $x \in \mathbb{R}^{H \times W \times C}$, the backbone maps it to a high-level feature vector $f_{\text{ext}} = \phi(x) \in \mathbb{R}^d$, with $d$ denoting the dimensionality of the extracted features. During training, the backbone parameters are frozen to leverage robust, transferable visual features learned from ImageNet. These features capture low- and mid-level patterns, such as edges, textures and shapes, which generalize well to medical images. Freezing the backbone also plausibly reduces susceptibility to the Barren Plateau problem \cite{mcclean2018barren}, where gradients vanish exponentially with circuit depth, by simplifying the optimization landscape of the hybrid quantum-classical network. We do not measure this directly.

\subsection{Feature Compression}
The extracted feature vector ${f}_{\rm ext}$ is too high-dimensional ($d = 768$ for ViT-B/16 \cite{Dosovitskiy2020vitb16}) for quantum processing, as current hardware supports at most $\sim$100 noisy qubits and the frozen backbone cannot be modified to reduce it. We therefore project $f_{\rm ext}$ into $\mathbb{R}^N$ with $N \ll d$ using a trainable linear layer $W_c \in \mathbb{R}^{N \times d}$ with bias $b_c \in \mathbb{R}^N$. Forcing highly expressive ViT features through a low-dimensional bottleneck restricts the hypothesis space. This limits the extent to which the classifier can fit noise or spurious correlations in the medical imagery \cite{zhang2017}. The bottleneck is a structural prerequisite of the quantum head. Since the circuit can accept no more inputs than it has qubits and it constrains both the hybrid model and the classical control of Section~\ref{sec:isolation} equally, what the circuit adds within that constraint is characterised in Section~\ref{sec:theory}. An element-wise arctangent then produces the compressed representation
\begin{equation}
z=\arctan(W_{c}f_{\mathrm{ext}}+b_{c})\in
\left(-\tfrac{\pi}{2},\tfrac{\pi}{2}\right)^{N}.
\label{eq:bottleneck}
\end{equation} This serves as the input to the VQC.

\subsection{Variational Quantum Circuit}
The quantum module consists of a parameterized quantum circuit acting on $N$ qubits as shown in Fig.~\ref{fig:architecture}.
The compressed feature vector $z$ is encoded in the first VQC layer using \textit{angle encoding}. The element-wise arctangent activation constrains each component of $z$ to the range $(-\pi/2, \pi/2)$, avoiding ambiguities from the periodicity of the quantum rotation gates and ensuring a unique mapping to a quantum state \cite{hirai2024practical}. The vector $z$ is then mapped to a corresponding quantum state $\ket{\psi({z})}$ as
\begin{equation}
    |\psi({z})\rangle = \bigotimes_{j=1}^{N} R_Y(z_j)\,|0\rangle.
    \label{eq:encoding}
\end{equation}
Here, $\otimes$ denotes the tensor product, combining individual qubits into a joint $N$-qubit system \cite{nielsen2010quantum} and $|0\rangle = [1, 0]^\top$ represents a qubit in the zero state. The single-qubit rotation about the $y$-axis is defined as
\begin{equation}
R_Y(\theta) \coloneqq
\begin{pmatrix}
\cos\left(\frac{\theta}{2}\right) & -\sin\left(\frac{\theta}{2}\right) \\
\sin\left(\frac{\theta}{2}\right) & \cos\left(\frac{\theta}{2}\right)
\end{pmatrix}.
\end{equation}
We use $R_Y(\theta)$ since it maps real-valued features directly onto the quantum state amplitudes, whereas rotation about the $z$-axis $R_Z(\theta)\ket{0} = e^{-i\theta/2}\ket{0}$ affects only the relative phase.

Following angle encoding, the circuit employs an ansatz \cite{peruzzo2014variational} composed of multiple repeated layers, each consisting of parameterized rotational gates $R_Z(\cdot)$ and $R_Y(\cdot)$ and a ring of controlled-NOT (CNOT) gates. After passing through the parameterized circuit, the state evolves to $|\psi({\theta}, {z})\rangle$, the final state vector that depends on the input features $z$ and the trainable circuit parameters $\theta$.

To extract informative classical features from the final quantum state
$\ket{\psi(\boldsymbol{\theta}, \mathbf{z})}$, we measure the expectation values of the Pauli observables $\sigma_X$ and $\sigma_Z$, the standard measurements along the $x$- and $z$-axes of the Bloch sphere. Let $\sigma_{\alpha}^{(j)}$ with $\alpha \in \{X, Z\}$ be the Pauli operator acting on the $j$-th qubit \cite{nielsen2010quantum}. The VQC output vector is then defined as
\begin{equation}
y_{q}\coloneqq\big[\langle\sigma_{Z}^{(1)}\rangle,\dots,
\langle\sigma_{Z}^{(N)}\rangle,
\langle\sigma_{X}^{(1)}\rangle,\dots,
\langle\sigma_{X}^{(N)}\rangle\big]^{\top}\in\mathbb{R}^{2N},
\label{eq:readout}
\end{equation}
where
\begin{equation}
\langle\sigma_{\alpha}^{(j)}\rangle
=
\langle
\psi(\boldsymbol{\theta},\mathbf{z})
|
\sigma_{\alpha}^{(j)}
|
\psi(\boldsymbol{\theta},\mathbf{z})
\rangle
\in
\mathbb{R}
\end{equation}is the quantum expectation value with $\langle \psi({\theta}, {z}) |$ denoting the conjugate transpose of $\ket{\psi({\theta}, {z})}$. These expectation values map the complex-valued quantum state to a real-valued feature vector suitable for a classical classifier.

\subsection{Classification }
The feature vector ${{y}_q}$ is fed into a fully-connected single-layer neural network followed by a softmax activation. This produces the class probability vector $y \in \mathbb{R}^c$. The overall hybrid model is trained end-to-end, excluding the feature extractor, by minimizing the cross-entropy loss \cite{mari2020transfer} with respect to the parameters of the classifier and quantum circuit.

\subsection{Function Class of a Single-Encoding Quantum Head}
\label{sec:theory}

Having specified the architecture, we now ask what the quantum head can express. The question matters because it determines, independently of any experiment, how sharply the decision of the model can vary between two images the backbone maps to nearby codes. Label corruption produces exactly such pairs, visually similar images carrying conflicting labels.

\medskip
\noindent\textbf{Restricted frequency support.}
Consider any Pauli observable $\sigma$ and the corresponding expectation $f_{\sigma}(z)=\langle\psi(\boldsymbol{\theta},z)|\,U^{\dagger}(\boldsymbol{\theta})\,\sigma\,U(\boldsymbol{\theta})\,|\psi(\boldsymbol{\theta},z)\rangle$, where the ansatz $U(\boldsymbol{\theta})$ is independent of $z$. Because the state \eqref{eq:encoding} is a product of single-qubit rotations, its amplitudes are products of $\cos(z_{j}/2)$ and $\sin(z_{j}/2)$ terms and the expectation is quadratic in those amplitudes. Every occurrence of $z_{j}$ therefore appears through one of $\cos^{2}\!\tfrac{z_{j}}{2}=\tfrac{1+\cos z_{j}}{2}, 
\sin^{2}\!\tfrac{z_{j}}{2}=\tfrac{1-\cos z_{j}}{2}$, or
$\sin\tfrac{z_{j}}{2}\cos\tfrac{z_{j}}{2}=\tfrac{\sin z_{j}}{2},$
so the frequency support of $f_{\sigma}$ in each coordinate is confined to $\Omega_{j}=\{-1,0,+1\}$. Each output is consequently a \emph{multilinear} trigonometric polynomial,
\begin{equation}
f_{\sigma}(z)=\sum_{S\subseteq\{1,\dots,N\}}\;
\sum_{\tau\in\{\sin,\cos\}^{|S|}}
a_{S,\tau}\prod_{j\in S}\tau_{j}(z_{j}),
\label{eq:fourier}
\end{equation}
of degree at most one in each variable. Only the coefficients $a_{S,\tau}$ depend on $\boldsymbol{\theta}$. Training reweights a fixed low-frequency basis and cannot introduce higher harmonics. This is the single-encoding case of the general result of \citeauthor{schuld2021} \shortcite{schuld2021}, which shows that the accessible spectrum is determined by the eigenvalues of the encoding generators and grows only when data is re-uploaded.

\medskip
\noindent\textbf{Bounded Jacobian.}
Because $\sigma$ has eigenvalues $\pm1$ we have $|f_{\sigma}|\le1$ and since $z_{j}$ enters through a single rotation generator the parameter-shift identity \cite{mitarai2018,schuld2019} gives
\begin{equation}
\frac{\partial f_{\sigma}}{\partial z_{j}}
=\frac{1}{2}\Big[f_{\sigma}\big(z+\tfrac{\pi}{2}e_{j}\big)
-f_{\sigma}\big(z-\tfrac{\pi}{2}e_{j}\big)\Big]
\;\Longrightarrow\;
\left|\frac{\partial f_{\sigma}}{\partial z_{j}}\right|\le1 .
\label{eq:lipschitz}
\end{equation}
The head maps the $N$-dimensional code $z$ to the $2N$ measurement outputs of \eqref{eq:readout}, so its Jacobian $J\in\mathbb{R}^{2N\times N}$ has $2N$ rows and $N$ columns. Every one of its $2N^{2}$ entries is bounded by unity \emph{uniformly in the trainable parameters}, so $\|J\|_{F}\le N\sqrt{2}$, a Lipschitz constant independent of $\theta$.

Equations~\eqref{eq:fourier} and~\eqref{eq:lipschitz} jointly bound the frequency content and local sensitivity available to fit exactly this kind of corruption.

The contrast with the classical control of Section~\ref{sec:isolation} is a contrast in the \emph{shape} of capacity rather than its quantity. An $N\rightarrow2N$ ReLU layer with weights $A_{1},A_{2}$, distinct from the compression matrix $W_{c}$ of \eqref{eq:bottleneck}, has Lipschitz constant $\|A_{2}\|\|A_{1}\|$, which is unbounded and grows freely during training and its piecewise-linear structure admits arbitrarily many decision regions within $(-\pi/2,\pi/2)^{N}$. At matched parameter count the two heads therefore differ in how sharply they can respond to small displacements in the code, not in how many parameters they carry.

We emphasise what this argument does \emph{not} establish. It does not show the head is incapable of fitting corrupted labels. It still operates on an $N$-dimensional code and the compression layer preceding it is trainable, so gradient descent can reposition examples within the bottleneck such that even a smooth head separates them. The prediction is narrower, that the two heads differ not in \emph{whether} they can fit corruption but in how readily and stably they do so. Section~\ref{sec:dynamics} tests this and also verifies the bound above by measuring the trained head's sensitivity to small displacements of the code.

Two further predictions follow. Since circuit depth reweights \eqref{eq:fourier} without extending its support, accuracy should saturate rather than improve monotonically with $L$, which the depth ablation in Section~\ref{sec:ablations} confirms. And since dual-basis readout adds independent low-frequency projections rather than higher harmonics, it should improve clean accuracy without compromising robustness.

\subsection{Trainable Parameter Count}

The hybrid architecture has $\sim$6,400 trainable parameters (compression layer: 6,152; VQC: 64; classifier: 153), about 1\% quantum, far below ResNet-18 ($\sim$10M) or fine-tuned ViT-B/16 ($\sim$86M). We do not claim parameter efficiency as a contribution; Section~\ref{sec:isolation} shows a comparably-sized classical head matches clean-label performance.

\section{Experiments and Results}

\subsection{Datasets and Evaluation Metrics}

We evaluate on six MedMNISTv2 multi-class datasets \cite{medmnistv2}: PathMNIST, BloodMNIST, DermaMNIST, OCTMNIST, OrganAMNIST and OrganCMNIST. These span $4$ to $11$ classes and over $325{,}000$ clinical-grade images, from roughly $10{,}000$ in DermaMNIST to $109{,}000$ in OCTMNIST. All images are resized to $224\times224\times3$. We report classification accuracy and because several datasets are strongly imbalanced, we also report AUROC, macro F1 and balanced accuracy briefly. Results are the mean and standard deviation over three runs with independent random initializations of the circuit parameters.

\subsection{Implementation}
\label{sec:impl}

Unless stated otherwise we use a frozen ImageNet-pretrained ViT-B/16 backbone, a circuit with $N=8$ qubits and $L=4$ layers and dual Pauli-$X$/Pauli-$Z$ readout. All the models are trained for up to 30 epochs with the Adam optimizer and a batch size of 32. We set the learning rates to $\eta_{q}=0.01$ for the VQC parameters and $\eta_{c}=0.003$ for the classical compressor and classifier. To prevent overfitting, we apply a 0.2 dropout in early classical layers and an early stopping strategy based on validation loss with a 5-epoch patience. Experiments were run on an Intel Xeon CPU using PennyLane's lightning.qubit simulator \cite{kang2025pennylane,2018arXiv181104968B}. Gradient computation for the quantum circuit was performed using the Adjoint differentiation method to ensure memory efficiency.

\subsection{Comparison with Hybrid Baselines Under Clean Labels}
\label{sec:clean}

Table~\ref{tab:hybrid} compares our model with state-of-the-art quantum-classical hybrids under clean labels. The proposed method leads on five of the six benchmarks, improving on the strongest prior hybrid (NQNN) by 17.4 and 15.7 percentage points on BloodMNIST and PathMNIST. Table~\ref{tab:symmetric} sources the same baselines; for both, full parity in training budget, backbone and hyperparameter search with our setup is not established. Performance is retained as the class count grows: on the eleven-class OrganAMNIST and OrganCMNIST, the model reaches $88.3\%$ and $85.7\%$, whereas QTNet and HQNet collapse to approximately $12\%$, near the eleven-class random baseline, a failure pattern plausibly consistent with barren plateaus in fully parameterized circuits \cite{qtnet2024dhara,mcclean2018barren}, though we do not measure gradients directly. Freezing the backbone and confining the trainable quantum parameters to $64$ keeps our optimization problem in a regime where this does not arise. NQNN remains stronger on DermaMNIST ($82.7\%$ against $75.2\%$), a dataset on which our model is weakest overall; we return to this in Section~\ref{sec:ablations}.

\begin{table*}[t]
\centering

\fontsize{9}{10}\selectfont
    \begin{tabular}{ c c c c c c c c }
    \hline
    \noalign{\vskip .5pt}
    Classes & Dataset & \textbf{Ours} & NQNN~\shortcite{rahman2025nqnn} & QCNN~\shortcite{qcnn2022bokhan} & QTNet~\shortcite{qtnet2024dhara} & HQNet~\shortcite{hqnet2024mazher} & A-HQCNN~\shortcite{ahqcnn2023ajlouni} \\ \noalign{\vskip 1.5pt}
    \hline
    \noalign{\vskip .5pt}
    4  & OCTMNIST    & \textbf{69.4} & 59.1          & 38.5 & 25.8 & 24.9 & 24.8 \\ \noalign{\vskip .5pt}
    7  & DermaMNIST  & 75.2  & \textbf{82.7} & 41.5 & 20.7 & 32.6 & 25.4 \\ \noalign{\vskip .5pt}
    8  & BloodMNIST  & \textbf{91.2} & 73.7          & 62.5 & 12.3 & 12.3 & 20.2 \\ \noalign{\vskip .5pt}
    9  & PathMNIST   & \textbf{90.9} & 75.2          & 61.7 & 14.0 & 11.6 & 12.7 \\ \noalign{\vskip .5pt}
    11 & OrganAMNIST & \textbf{88.3} & 81.4          & 61.8 & 12.5 & 12.4 & 12.0 \\ \noalign{\vskip .5pt}
    11 & OrganCMNIST & \textbf{85.7} & 79.9          & 63.0 & 13.2 & 12.0 & 21.7 \\ \noalign{\vskip .5pt}
    \hline
    \end{tabular}
    \normalsize
\caption{Classification accuracy (\%) under clean labels against quantum-classical hybrid models. Baseline results are sourced from \cite{rahman2025nqnn}. All results in this table use the ViT-B/16 backbone with $L=4$ and Pauli $X$+$Z$ readout. Bold indicates the best result per dataset.}
\label{tab:hybrid}
\end{table*}

\subsection{Robustness to Label Noise}
\label{sec:noise}
We now turn to the central results. We consider two corruption models. Symmetric noise flips labels uniformly at random, independently of image content, simulating unstructured annotation error. Instance-dependent noise (IDN) makes the flip probability a function of the visual ambiguity of each image and is the more realistic model of clinical annotation error.

\paragraph{Symmetric noise.}
Table~\ref{tab:symmetric} reports accuracy at $10\%$, $30\%$ and $50\%$ symmetric noise against DUE-Net \cite{duenet}, RWNet \cite{diaz2021} and NQNN. At $10\%$, our model is strongest on four of six datasets, with NQNN ahead on DermaMNIST and DUE-Net on OrganCMNIST. From $30\%$ upward, it leads on all six and the margins widen with the noise rate. At $50\%$ we retain $89.8\%$ on PathMNIST where the best baseline reaches $58.0\%$ and $82.3\%$ on OrganAMNIST against $61.7\%$. The pattern is the signature of the mechanism proposed in Section~\ref{sec:theory}. Methods that rely on capacity plus explicit correction degrade steeply once the corruption exceeds what their correction can estimate, whereas a model that structurally cannot express the corruption degrades only as fast as the true signal is diluted.

\begin{table*}[t]
\centering
\fontsize{9}{10}\selectfont
\begin{tabular}{c  cccc  cccc  cccc}
\hline \noalign{\vskip 0.5pt}
\textbf{NR} & \multicolumn{4}{c}{\textbf{10\%}} & \multicolumn{4}{c}{\textbf{30\%}} & \multicolumn{4}{ c}{\textbf{50\%}} \\ \noalign{\vskip 0.5pt}
\hline \noalign{\vskip 0.5pt}
\textbf{Dataset} & DUE & RW &  NQNN & \textbf{Ours} & DUE & RW & NQNN & \textbf{Ours} & DUE & RW & NQNN & \textbf{Ours} \\ \noalign{\vskip 0.5pt}
\hline \noalign{\vskip 0.5pt}
OCT    & 52.4 & 50.0 & 55.3 & \textbf{65.0} & 45.5 & 44.3 & 46.6 & \textbf{64.2} & 40.6 & 40.7 & 41.7 & \textbf{58.0} \\ \noalign{\vskip 0.5pt}
Derma  & 41.6 & 23.0 & \textbf{80.3} & 73.9 & 28.0 & 22.4 & 71.0 & \textbf{72.2} & 24.2 & 19.3 & 66.1 & \textbf{69.3} \\ \noalign{\vskip 0.5pt}
Blood  & 82.3 & 79.7 & 70.4 & \textbf{89.8} & 64.6 & 54.2 & 64.8 & \textbf{86.3} & 49.8 & 43.9 & 61.9 & \textbf{80.0} \\ \noalign{\vskip 0.5pt}
Path   & 80.2 & 76.2 & 71.3 & \textbf{90.8} & 59.1 & 60.1 & 63.5 & \textbf{90.2} & 48.4 & 44.6 & 58.0 & \textbf{89.8} \\ \noalign{\vskip 0.5pt}
OrganA & 29.0 & 44.8 & 80.3 & \textbf{87.2} & 28.9 & 35.0 & 73.9 & \textbf{85.0} & 23.9 & 28.3 & 61.7 & \textbf{82.3} \\ \noalign{\vskip 0.5pt}
OrganC & \textbf{86.9} & 83.5 & 78.3 & 82.5 & 68.7 & 65.0 & 59.9 & \textbf{79.5} & 48.8 & 46.9 & 61.6 & \textbf{74.9} \\ \noalign{\vskip 0.5pt}
\hline
\end{tabular}
\normalsize
\caption{Classification accuracy (\%) comparison under different symmetric label noise ratios (NR). Baseline results are sourced from \cite{rahman2025nqnn}. Bold values indicate the highest accuracy for each noise level.}
\label{tab:symmetric}
\end{table*}

\paragraph{Instance-dependent noise.}
Following Algorithm~1 of \cite{ma2025lnmbench} we generate feature-dependent label transitions using an auxiliary ResNet-18 \cite{kaiming2016resnet} to estimate per-instance difficulty and compare against the full classical LNMBench (CE-V) pipeline \cite{ma2025lnmbench} on Path, Blood, OrganC and Derma. At $20\%$ IDN, LNMBench leads on all four by $3.9$ to $7.7$ points and our model is competitive but not leading. At $50\%$ the ordering partially reverses. We improve on PathMNIST ($79.3\%$ against $76.3\%$), match within one standard deviation on BloodMNIST and DermaMNIST and remain behind on OrganCMNIST. The trend matches the symmetric-noise results, in that our relative position improves as corruption intensifies, but we do not claim superiority under IDN. This is the expected limit of the mechanism, instance-dependent flips concentrate on genuinely ambiguous cases, so the corruption is partly \emph{aligned} with the smooth class structure and therefore, unlike symmetric noise, partly within the frequency band the head can express.

\subsection{Isolating the Quantum Module}
\label{sec:isolation}

To attribute the robustness to the quantum head rather than to any other component, we construct a control that differs from the full model in exactly one respect. It shares the frozen ViT-B/16 backbone, the $768\rightarrow8$ compression layer, the arctangent activation, the dropout, the $16$-dimensional intermediate width, the classifier, the optimiser and the learning rate but the only change is that the
$8\rightarrow16$ map is a ReLU layer rather than the circuit. Trainable parameter counts are $6{,}369$ for the hybrid model and $6{,}449$ for the control on PathMNIST.

Table~\ref{tab:isolation} shows the result. Under clean labels the two are indistinguishable - $90.94\%$ against $90.92\%$ on PathMNIST and within $1.9$ points on every dataset, with the control marginally ahead on OrganAMNIST and DermaMNIST. The restricted head therefore costs nothing in discriminative power and no claim of clean-label superiority over classical architectures is warranted.

The behaviour under corruption is entirely different. The gap grows monotonically with the noise rate on every dataset. At $50\%$ symmetric noise the hybrid model improves on the control on five of six datasets, by $10.8$ points on PathMNIST ($89.75\%$ against $78.96\%$), $8.2$ on OCTMNIST, $7.0$ on OrganAMNIST, $3.3$ on OrganCMNIST and $2.6$ on BloodMNIST; on DermaMNIST the two are equal to within $0.03$ points. PathMNIST is the most striking case. The hybrid model loses $1.2\%$ relative to its own clean-label accuracy while half its training labels are wrong, whereas the control loses $12.0\%$.

This comparison controls for trainable parameter count and shared upstream representation while deliberately varying inductive bias and expressivity geometry. Because the learning rate and schedule are shared, the gap cannot be explained by differing optimization speed. We attribute it to the difference in function class established in Section~\ref{sec:theory}, the circuit restricts the model to smooth, low-frequency decision boundaries, whereas the ReLU head can sharpen its response arbitrarily as training proceeds. Section~\ref{sec:dynamics} examines this attribution directly and refines it, showing that the difference lies in stability under prolonged training rather than in an outright inability to fit corrupted targets.

\begin{table*}[t]
\centering
\fontsize{9}{10}\selectfont
 \begin{tabular}{c  l  c  c  c  c }
\hline \noalign{\vskip 0.5pt}
\textbf{Dataset} & \textbf{Model} & \textbf{Ideal} & \textbf{10\% Noise} & \textbf{30\% Noise} & \textbf{50\% Noise} \\ \noalign{\vskip 0.5pt}
\hline \noalign{\vskip 0.5pt}
\multirow{2}{*}{PathMNIST}& Classical  & 90.92 & 89.50  & 86.56 & 78.96  \\ \noalign{\vskip 0.5pt}
 & Ours & \textbf{90.94}  & \textbf{90.82} & \textbf{90.16}  & \textbf{89.75} \\ \noalign{\vskip 0.5pt}
\hline \noalign{\vskip 0.5pt}
\multirow{2}{*}{OCTMNIST} & Classical  & 65.50  & 60.47 & 56.63 & 49.80 \\ \noalign{\vskip 0.5pt}
& Ours & \textbf{65.57}  & \textbf{65.03}  & \textbf{64.23} & \textbf{57.97} \\ \noalign{\vskip 0.5pt}
\hline \noalign{\vskip 0.5pt}
\multirow{2}{*}{OrganAMNIST} & Classical & \textbf{89.08}  & \textbf{87.29}  & 82.91 & 75.33 \\ \noalign{\vskip 0.5pt}
& Ours & 88.27 & 87.15  & \textbf{85.04} & \textbf{82.34} \\ \noalign{\vskip 0.5pt}
\hline \noalign{\vskip 0.5pt}
\multirow{2}{*}{DermaMNIST} & Classical & \textbf{75.61} & \textbf{74.49} & 71.59 & \textbf{69.34} \\ \noalign{\vskip 0.5pt}
& Ours & 75.23 & 73.90 & \textbf{72.22} & 69.31 \\ \noalign{\vskip 0.5pt}
\hline \noalign{\vskip 0.5pt}
\multirow{2}{*}{BloodMNIST} & Classical & 91.09 & 89.46 & 85.65 & 77.43 \\ \noalign{\vskip 0.5pt}
& Ours & \textbf{91.17} & \textbf{89.76} & \textbf{86.31} & \textbf{80.03} \\ \noalign{\vskip 0.5pt}
\hline \noalign{\vskip 0.5pt}
\multirow{2}{*}{OrganCMNIST} & Classical & 83.80 & \textbf{83.92} & \textbf{79.70} & 71.57 \\ \noalign{\vskip 0.5pt}
& Ours & \textbf{85.68} & 82.46 & 79.53 & \textbf{74.85} \\ \noalign{\vskip 0.5pt}
\hline
\end{tabular}
\normalsize
\caption{Hybrid VQC head against a classical ReLU head that shares the same frozen backbone, $768\rightarrow8$ bottleneck, intermediate width, classifier, optimizer and learning rate and differs only in that the $8\rightarrow16$ map is a ReLU layer rather than the circuit. Symmetric label noise; accuracy (\%).}
\label{tab:isolation}
\end{table*}

\subsection{Memorization Dynamics}
\label{sec:dynamics}
Table~\ref{tab:isolation} measures a downstream consequence of reduced memorization; here we measure memorization itself. Training both heads on PathMNIST at $50\%$ symmetric noise and retaining the corruption mask, we track three quantities each epoch. On the flipped subset we measure agreement with the corrupted labels, which is memorization proper and with the \emph{true} labels, which is resistance to it. On the uncorrupted subset we measure fit to genuine signal, controlling for a head that simply learns more slowly. Both heads train for $60$ epochs under a shared learning rate with early stopping disabled, so late-stage memorization is visible rather than truncated away. For tractability the experiment uses a class-stratified subset of $10{,}000$ training images, so absolute accuracies fall below the full-data values of Table~\ref{tab:isolation}; the comparison between heads is unaffected, since both see identical data.

\medskip

The heads differ in stability, not capacity (Table~\ref{tab:memorization}). Corrupted- and clean-label agreement are near-identical at the end of training, ruling out the possibility that the quantum head merely optimizes more slowly; the difference emerges after the peak, where the classical head loses more than twice as much clean accuracy by epoch 60. A capacity test in the sense of \citeauthor{zhang2017} \shortcite{zhang2017}, retraining both heads on entirely randomized labels, confirms the interpretation: the quantum head reaches roughly 89\% of the classical head’s excess-over-chance memorization, indicating the shared eight-dimensional bottleneck, not the head, is the binding capacity constraint. The restricted frequency support therefore does not confer an inability to fit corrupted labels; it confers stability while doing so, which matters because a model that resists corruption only if stopped at the right epoch needs validation labels that can be trusted, precisely what is missing when annotations are unreliable.

The bound in Section~\ref{sec:theory} also holds empirically: perturbing the bottleneck code by a small random displacement at 4,000 training points and recording the ratio of output to input change gives sensitivity ratios two to four times lower for the quantum head than the classical one (Table~\ref{tab:memorization}), comfortably within $\|J\|_{F}\le N\sqrt{2}\approx11.3$. These are local, directional estimates and thus lower bounds on the true Lipschitz constant, which makes the separation conservative.

\begin{table}[ht]
\centering
\small
\begin{tabular}{p{3.2cm}cc}
\hline
\textbf{Metric} & \textbf{Quantum head} & \textbf{Classical head} \\
\hline
Corrupted-label agreement (epoch 60) & 7.0\% & 7.4\% \\
Clean-subset agreement (epoch 60) & 92.0\% & 92.5\% \\
Peak clean test accuracy (epoch) & 12 & 18 \\
Accuracy drop, peak $\to$ epoch 60 (pts) & 2.5 & 6.5 \\
Random-label agreement (capacity test) & 36.1\% & 39.1\% \\
Sensitivity ratio (random-label setting) & 0.71 & 2.89 \\
Sensitivity ratio (50\% noise setting) & 0.88 & 1.71 \\
\hline
\end{tabular}
\normalsize
\caption{Memorization dynamics on PathMNIST at 50\% symmetric noise (60 epochs, early stopping disabled). Chance-level agreement is 11.1\% (nine classes). Sensitivity ratio is the empirical output-change/input-change ratio under random bottleneck-code perturbation (Section~\ref{sec:theory}), a local lower bound on the Lipschitz constant.}

\label{tab:memorization}
\end{table}

\subsection{Ablations}
\label{sec:ablations}
The remaining design choices were selected by ablation; we summarise the outcomes rather than tabulate them.

\paragraph{Measurement basis.} Dual Pauli-$X$/Pauli-$Z$ readout is best on five of six datasets (91.2\% vs. 86.2\%/90.8\% single-basis on BloodMNIST); it adds independent low-frequency projections rather than higher harmonics, so it does not extend the frequency support of Section~\ref{sec:theory}.

\paragraph{Circuit depth.} Varying $L$ from $2$ to $10$, accuracy peaks at $L=4$ on four of six datasets and saturates or drifts down thereafter; $L=2$ is consistently insufficient and the exceptions are OCTMNIST and OrganAMNIST, which peak at $L=6$ and $L=10$ by $2.5$ and $0.4$ points. This is the saturation predicted in Section~\ref{sec:theory}, because data is encoded once, additional ansatz layers reweight the fixed basis of \eqref{eq:fourier} without extending its support, so depth cannot buy expressivity beyond what a few layers already access. We adopt $L=4$ throughout. The ablation was run under clean labels; the interaction between depth and noise robustness is left to future work.

\paragraph{Feature extractor.} ViT-B/16 is strongest on five of six datasets, consistent with global self-attention capturing long-range structure \cite{chen2021transunet}; the exception is OCTMNIST, where EfficientNet-B0 leads (69.4\% vs. 65.6\%), suggesting a local-texture task favors convolutional inductive bias. All robustness experiments use ViT-B/16.

\paragraph{Class imbalance.} Because several datasets are imbalanced we also measured AUROC, macro F1 and balanced accuracy. Class separation is strong throughout, with AUROC above $0.98$ on the four largest datasets. DermaMNIST requires qualification, an AUROC of $0.906$ coexists with a macro F1 of $0.438$ and balanced accuracy of $0.447$, meaning the model ranks cases informatively but its argmax decisions are dominated by majority classes and rare lesion categories are largely missed. Class imbalance is therefore \emph{not} a form of label unreliability this architecture handles; the robustness we claim is specific to label noise.

\subsection{Limitations}
\label{sec:limitations}

Several constraints bound what we claim.

\emph{Noiseless simulation.} All quantum computation is noiseless state-vector simulation. Hardware execution introduces decoherence, gate error and shot noise; shot noise plausibly reinforces the smoothing we rely on, but coherent gate errors may distort the learned function in ways this study cannot predict.

\emph{Input resolution.} MedMNISTv2 images are distributed at $28\times28$ and upsampled to the backbone's $224\times224$ input, so we do not process natively high-resolution images. The architectural claim, that resolution is decoupled from qubit count because the circuit sees only an eight-dimensional code, is unaffected and the source resolution is retained for comparability with the baselines of Table~\ref{tab:hybrid}.

\emph{Scope and encoding specificity.} Section~\ref{sec:dynamics} shows the circuit reduces sensitivity to prolonged training on corrupted labels, not that it cannot memorize; the random-label test shows it can. The frequency bound is moreover specific to single-layer angle encoding and data re-uploading would widen the accessible spectrum and, we predict, erode the advantage. Finally, robustness is bought by restricting expressivity. DermaMNIST trails NQNN by $7.5$ points under clean labels and the architecture is not a general-purpose improvement.

\section{Conclusion and Future Work}

We have introduced a hybrid quantum-classical transfer learning architecture for medical image classification under label noise, in which a frozen ViT-B/16 backbone, an eight-dimensional bounded bottleneck and an 8-qubit VQC classify $224 \times 224 \times 3$ RGB images while only eight qubits participate in the decision. The central empirical finding is robustness to label corruption without any explicit noise-handling component. Across six MedMNISTv2 benchmarks the architecture outperforms published quantum and noise-robust classical baselines at $30\%$ and $50\%$ symmetric noise, retaining $89.8\%$ on PathMNIST at $50\%$ noise where the strongest baseline reaches $58.0\%$ and it scales to 11-class problems where fully parameterized hybrid circuits collapse toward chance under barren plateaus. We also characterised the source of that robustness through comparative simulations. %rather than asserting it. 
A control sharing the backbone, bottleneck, parameter count and optimizer and replacing only the circuit with a ReLU layer, matches the hybrid model on clean labels but degrades faster as corruption rises. Direct measurement then refines the explanation that the circuit does not prevent memorization, reaching $89\%$ of the classical head's memorization on randomized labels, which indicates the shared bottleneck is the binding capacity constraint. What the circuit contributes is an inductive bias toward smooth decision functions, with measured sensitivity two to four times below the matched classical head and well inside the analytical bound, yielding markedly more stable behaviour under prolonged training on corrupted labels. 

Future work includes testing the falsifiable prediction that data re-uploading erodes the effect, evaluating whether other restricted-spectrum heads reproduce it classically and establishing whether the advantage survives execution on NISQ hardware.

{\fontsize{9.0pt}{10.0pt}\selectfont
\bibliography{aaai2026}
}

\end{document}